\documentstyle[prl,aps,twocolumn]{revtex}
\include{epsf}
\epsfverbosetrue
\begin{document}
\twocolumn[\hsize\textwidth\columnwidth\hsize\csname @twocolumnfalse\endcsname

\title{Microscopic mechanism for mechanical polishing of diamond (110) surfaces}
\author{M. R. Jarvis$^{(1)}$, R. P\'erez$^{(2)}$, F. M. van Bouwelen$^{(3)}$, and M. C. Payne$^{(1)}$}
\address{$^{(1)}$ 
Theory of Condensed Matter, Cavendish Laboratory,
University of Cambridge, Madingley Road, Cambridge CB3 OHE, UK}
\address{$^{(2)}$ 
Departamento de F\'{\i}sica Te\'orica de la Materia Condensada,
Universidad Aut\'onoma de Madrid, E-28049 Madrid, Spain}
\address{$^{(3)}$ 
Applied Physics, Faculty of Science, University of Nijmegen, Toernooiveld 1, 6525 ED Nijmegen, Netherlands}
\date{\today}
\maketitle

\begin{abstract}
Mechanically induced degradation of diamond, as occurs during polishing, is studied using total--energy pseudopotential calculations.  The strong asymmetry in the rate of polishing between different directions on the diamond (110) surface is explained in terms of an atomistic mechanism for nano--groove formation.  The post--polishing surface morphology and the nature of the polishing residue predicted by this mechanism are consistent with experimental evidence.
\end{abstract}
]
\narrowtext

The extreme mechanical properties of diamond are of both practical and academic interest, and have been the subject of much recent research \cite{R1}.  Diamond, the hardest known natural material, is extremely wear resistant and conventional polishing requires the abrasive action of diamond powder to wear down the surface.  In spite of extensive experimental investigation, very little is known about the dominant microscopic mechanisms which occur during polishing, where the conditions are such that both mechanical and chemical processes can play a role.  In this Letter we describe the results of {\it ab--initio} quantum--mechanical simulations of particular wear processes on the (110) surface.  These simulations suggest that the atomistic mechanism of nano--grooving may be responsible for the main features of diamond polishing on this surface \cite{R2}.


The most striking feature of diamond polishing on the (110) surface is the strong asymmetry in the rates of material removal between different directions.  Ratios of greater than 10:1 between the $\langle001\rangle$ (soft) direction and the $\langle 1\bar10 \rangle$ (hard) direction are typically observed \cite{R3}.  Scanning tunneling microscopy studies \cite{R4} of the post--polishing surface morphology also show significant differences between the directions.  In the soft direction there are grooves which are flat bottomed on the atomic scale and which are absent in the hard direction.  For the soft direction, the absence of fractures or dislocations indicates that the regime in which polishing occurs is below the critical stress.  Furthermore the size of clusters in the residue, and the small amount of material removed per traversal in sliding indenter experiments \cite{R4}, suggest a mechanism in which only a few carbon atoms are removed in each tribological event.  In other respects the hard and soft polishing directions are more similar, for example, the increase in polishing rate in hydrogen and oxygen rich environments \cite{R5} and the weak temperature dependence of the rate \cite{R6}.  Studies of the polishing residue show both similarities and differences between the directions.  Electron energy loss spectroscopy \cite{R6} and transmission electron microscopy \cite{R7} studies of the polishing residue indicate that it consists mainly of amorphous clusters, although nanometer scale regions with a structure similar to graphite are also observed.  When polishing in the hard direction, additional small diamonds ($\sim$1 nm$^3$) are occasionally observed in the residue, although it is unclear whether these originated in the polishing diamond or the polished diamond, and there are also fewer graphitic regions.  


Existing theories of polishing, based on microcleavage \cite{R7.5}, are, as far as the soft direction is concerned, not consistent with this new experimental
evidence.  In the hard direction, this experimental evidence does not suggest cleavage along the (111) planes but does not fully rule out a different fracture mechanism.  As yet no detailed atomic mechanism explaining the polishing anisotropy has been proposed.


To investigate the asymmetry in the wear rate and the differences in post--polishing surface morphology we performed simulations in which a single nano--asperity, constructed on the (110) surface, was deformed by a rigid tip.  We first present the results of separate simulations in which the tip was incident from the soft and the hard polishing directions, which show pronounced differences in the extent of the induced deformation.  Then these differences in asperity removal are related to the process of nano--grooving which, we believe, is responsible for the polishing anisotropy.  Finally we present a simple schematic representation of the nano--grooving process which highlights the qualitative differences between the directions.  This wear process is predominantly mechanical rather than chemical in agreement with the suggestion of Couto {\it et al.} \cite{R8}.


The implementation of a standard total--energy pseudopotential method \cite{R9} on a massively parallel computer permits the quantum mechanical simulation (with an accurate description of bond breaking and formation) of systems containing several hundred atoms, sufficient for the modelling of a realistic nano--asperity and tip.  

\begin{figure}[htbp] 
\begin{center}
\hspace{2.5cm}
\epsfxsize=1.0cm \epsffile{./EPS/110.eps} 
\end{center}
\vspace{-3.0cm} 
\begin{center}
\hspace*{-0.8cm} \epsfxsize=5.0cm \epsffile{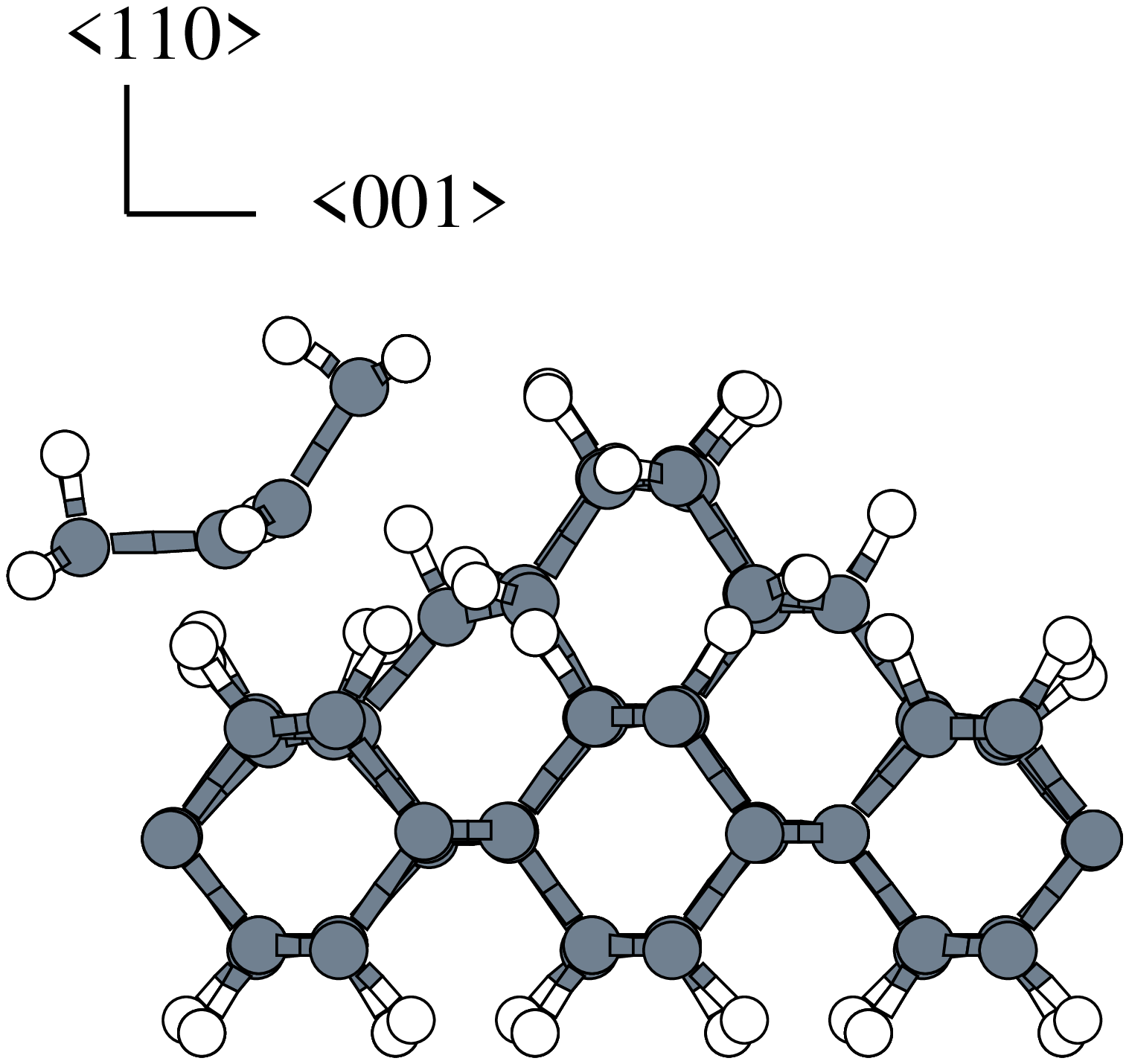} \hspace*{-0.8cm} \epsfxsize=5.0cm \epsffile{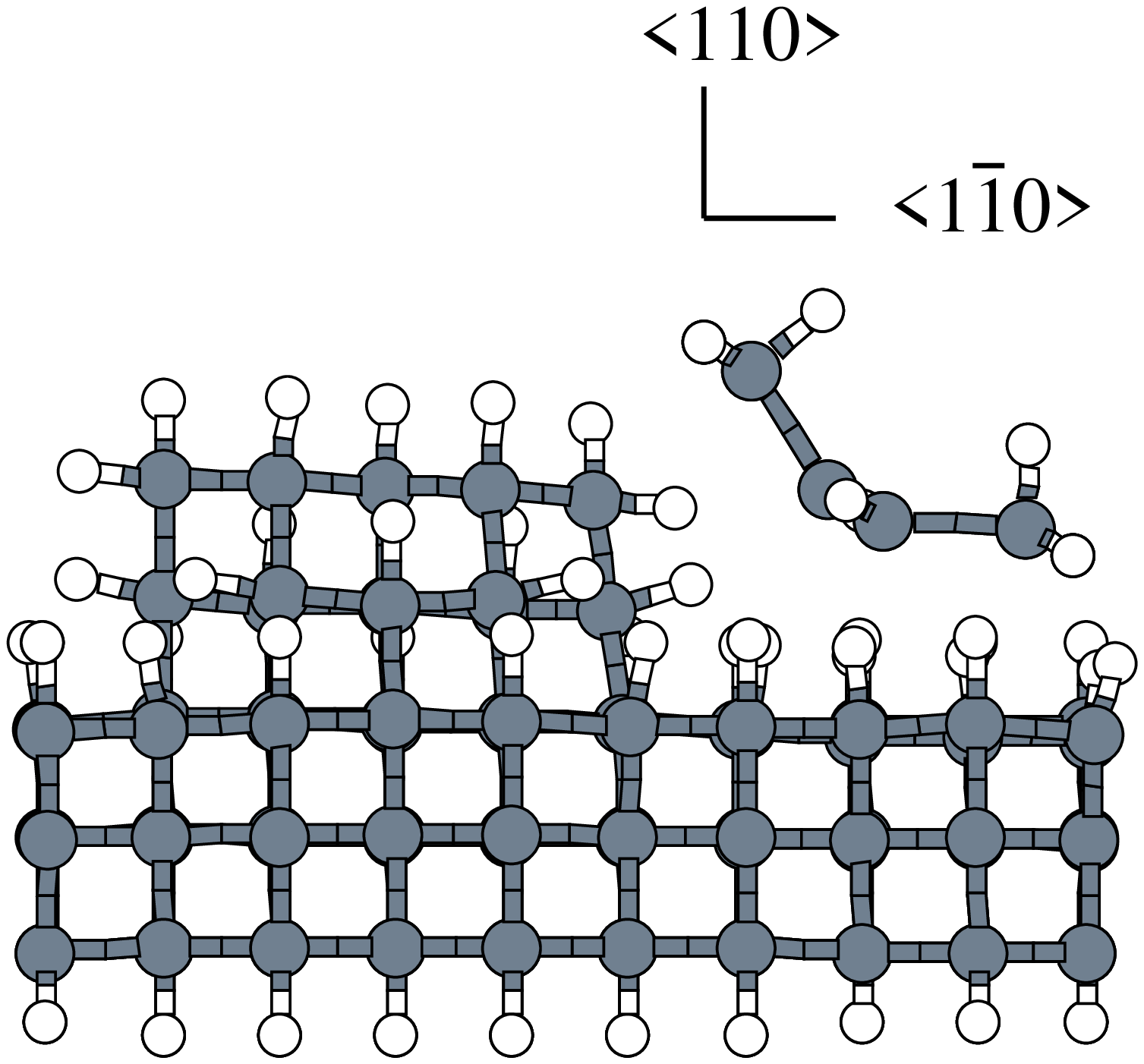} 
\end{center}
\vspace{-2.0cm}
\caption{Relaxed initial atomic positions: (Left) Soft direction simulation viewed along $\langle 1\bar10 \rangle$. This supercell has a volume of 1291 \AA$^3$ and contains 93 C and 68 H atoms. (Right) Hard direction simulation viewed along $\langle 001 \rangle$.  This supercell has a volume of 1603 \AA$^3$ and contains 111 C and 80 H atoms.  (Center)  Top view of the (110) surface showing the two scanned directions: $\langle 1\bar10 \rangle$ parallel to the characteristic chains and $\langle 001 \rangle$ perpendicular to them.}                          
\label{Initial_Position}                      
\end{figure}

The asperity was constructed by adding tetrahedrally bonded carbon atoms onto a clean, unreconstructed, (110) surface.  All dangling back--bonds were then saturated with hydrogen and the system was allowed to relax.  The base plane of carbons of this relaxed system was then held rigid during the subsequent simulations.  We chose to saturate the surface with hydrogen because of the large increase in polishing rate in hydrogen--rich environments and the known affinity of diamond surfaces for hydrogen \cite{R10}.  We formed a rigid tip, with which to deform the asperity, from a fragment of a Pandey $(2\times 1)$ reconstructed (111) surface \cite{R11,R11.5}.  The dangling C--C bonds of this fragment were saturated with hydrogen.  The (111) surface is realistically modelled as a rigid tip as this surface is the natural cleavage plane of diamond and is known to be relatively hard in all directions.  The Pandey reconstructed tip was chosen in preference to a hydrogen saturated (111) surface as, although polishing is performed at temperatures below the desorption temperature of hydrogen, the wear process will remove any hydrogen from the abrading diamond.  For both directions the rigid tip was advanced into the asperity in steps of 0.1 \AA.  At each position we allowed full relaxation of the remaining atomic degrees of freedom.  This adiabatic approximation is justified as the speed of sound in diamond is approximately three orders of magnitude greater than the typical velocity of the abrading diamond.  Figure \ref{Initial_Position} shows the initial atomic configurations of the supercells \cite{R15}.



The local density approximation was used to describe exchange and correlation, and the electronic states were expanded at the $\Gamma$ point of the Brillouin zone using a plane--wave basis set truncated at 44 Ry.  The calculation was converged with respect to plane wave cutoff and supercell height to an accuracy of 0.1eV per atom.  The electrostatic interaction between the asperities in neighbouring supercells was monitored during the simulations and was found to be insignificant.  An optimized non--local pseudopotential \cite{R12} was used to describe scattering from the 1s core of the carbon atoms and a Coulomb potential was used for hydrogen.  The non--local pseudopotential was applied in the Kleinman--Bylander form \cite{R13}.  

\begin{figure}[htbp]                   
\begin{center}
\vspace*{-1.2cm}                      
\epsfxsize=9.0cm                       
\hspace*{0.0cm}
\epsffile{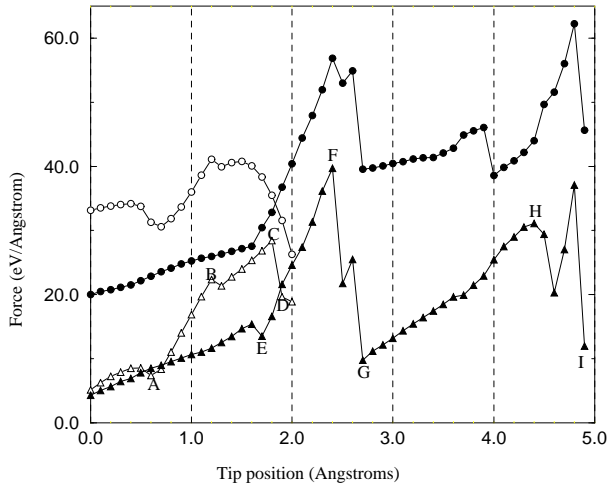}
\vspace*{-0.5cm}   
\end{center}                          
\caption{Forces on tip during the simulation.  Open symbols refer to the soft direction, filled to the hard direction.  Triangles refer to the retarding force, i.e. the force which opposes the advance of the tip, and circles refer to the normal force.} 
\label{Forces}                      
\end{figure}

The force felt by the tip as a function of its position (Figure \ref{Forces}) shows some obvious differences between the hard and soft directions.  In the hard direction, the tip can be advanced much further before reaching a yield point, beyond which all rigidity is lost.  Hence, since the forces on the tip are of comparable size for both directions, the impulse delivered to the tip is larger in the hard direction.  However, this is not sufficient to explain the anisotropy in the polishing rate.  Another feature of the hard direction is the large spike in the forces which might deflect the incident tip. 


The reason that the normal force is larger in the soft direction is that the tip was at a slightly lower initial height above the surface during this simulation.  The height and angle of the tip were chosen to make reasonable contact with the asperity for the simulations but this is not believed to be critical to the mechanism.  The justification for this assertion is evidence from simulations in which the rigid tip was replaced by a term in the Hamiltonian representing fictitious mechanical forces \cite{R18}.  These forces had no component normal to the surface and the asperity was allowed to relax as the forces were steadily increased.  The results of these simulations indicate that the process is predominantly mechanical and that the role of the normal force is not important to the mechanism in the soft direction although it does help to stabilize the asperity in the hard direction by preventing it from peeling off.  In view of this evidence we expect no qualitative change in the mechanism provided that the tip makes reasonable contact with the asperity but does not cause direct subsurface damage.  

\begin{figure}[htbp]                   
\begin{center} 
\hspace*{-0.4cm} \epsfxsize=5.0cm \epsffile{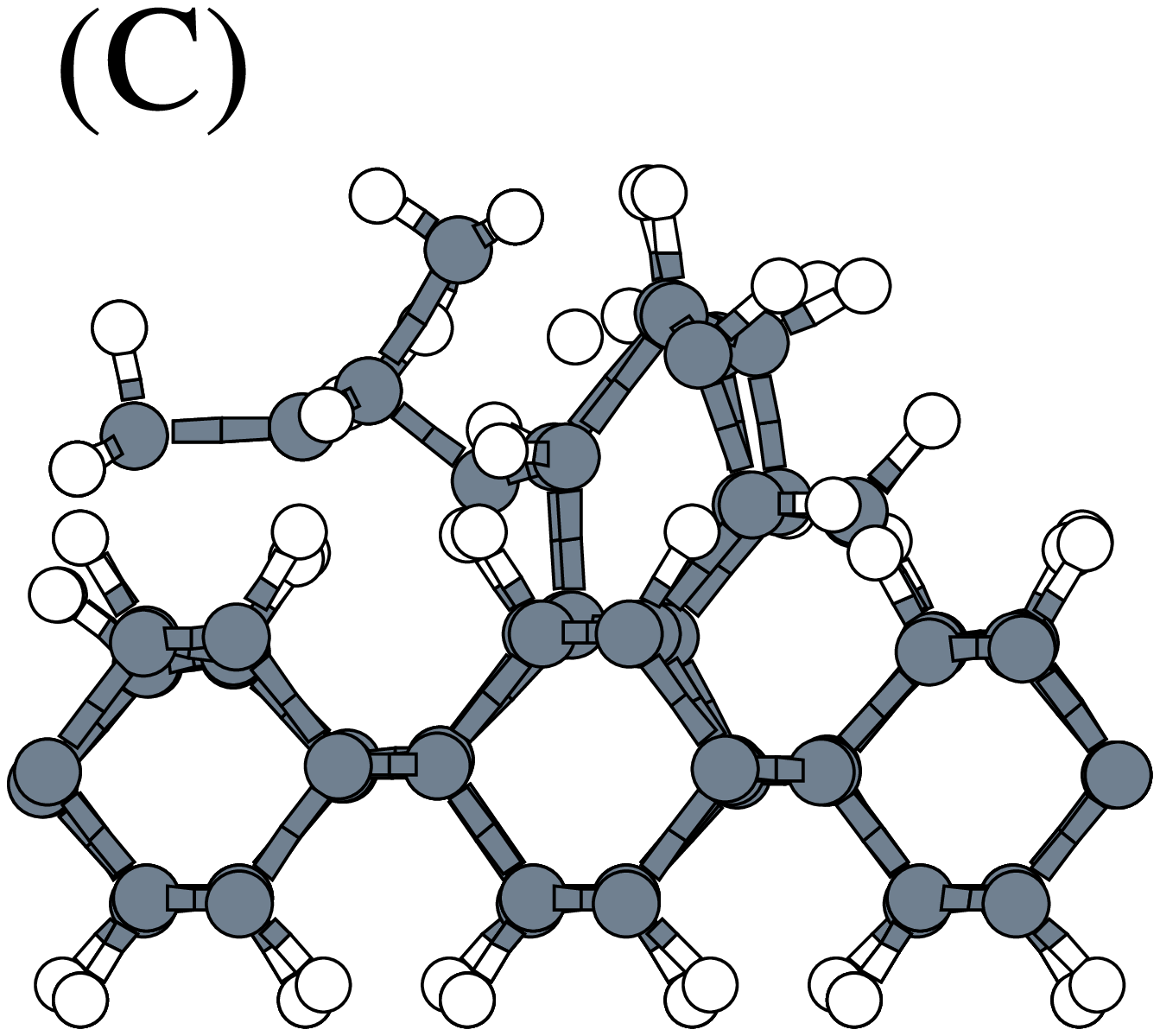} \hspace*{-1.2cm} \epsfxsize=5.0cm \epsffile{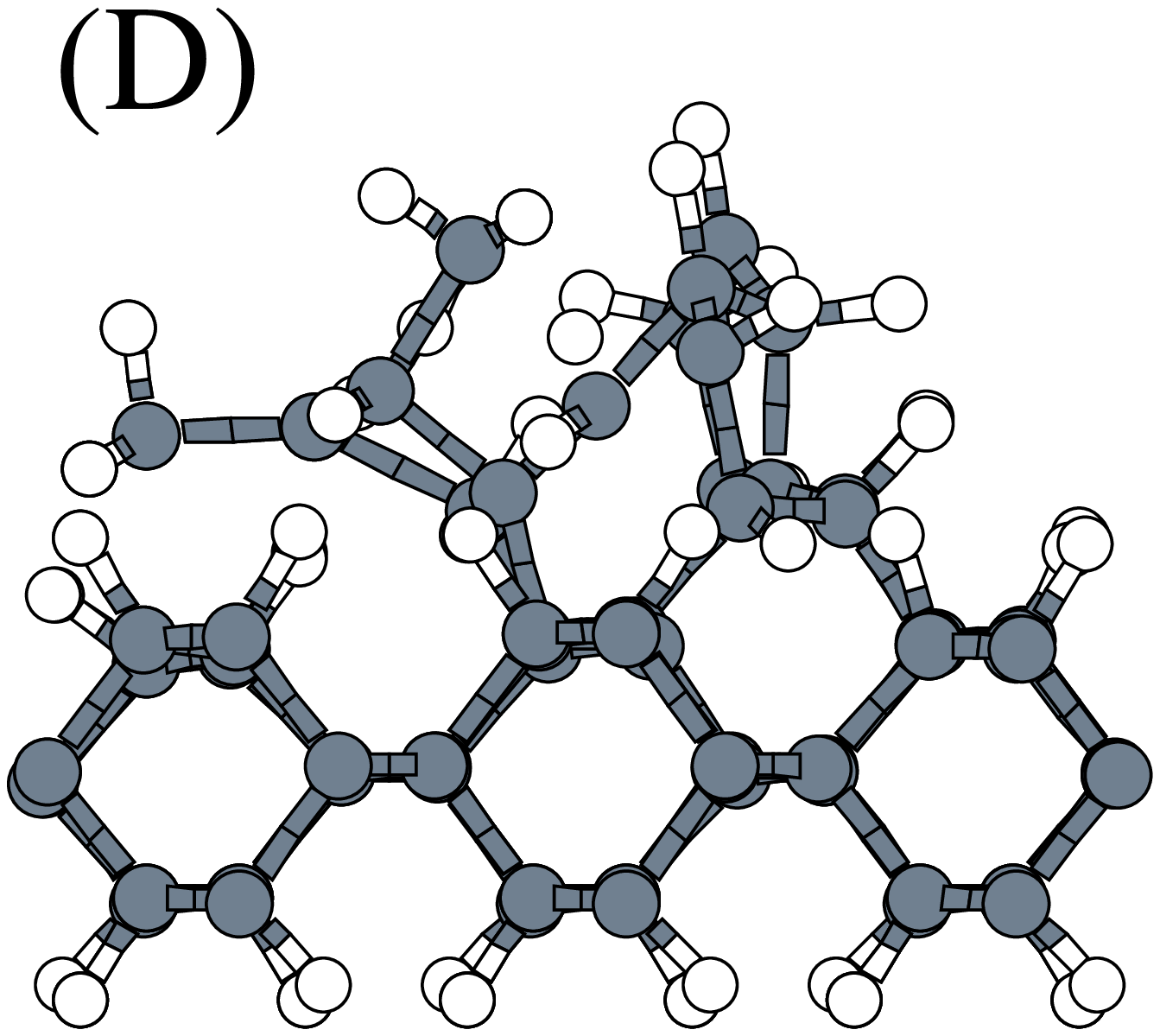}   
\vspace*{-2.0cm} 
\end{center}                           
\caption{Atomic configurations, corresponding to the points shown on the force graph (Figure \ref{Forces}), relating to the simulation performed in the soft polishing direction.  The diagrams correspond to tip positions (A) = 0.6 \AA, (B) = 1.2 \AA, (C) = 1.8 \AA \hspace*{0.1cm}and (D) = 1.9 \AA.}              
\label{Soft}                         
\end{figure}

\begin{figure}[htbp]                                                   
\begin{center}\hspace*{-1.00cm} \epsfxsize=5.0cm \epsffile{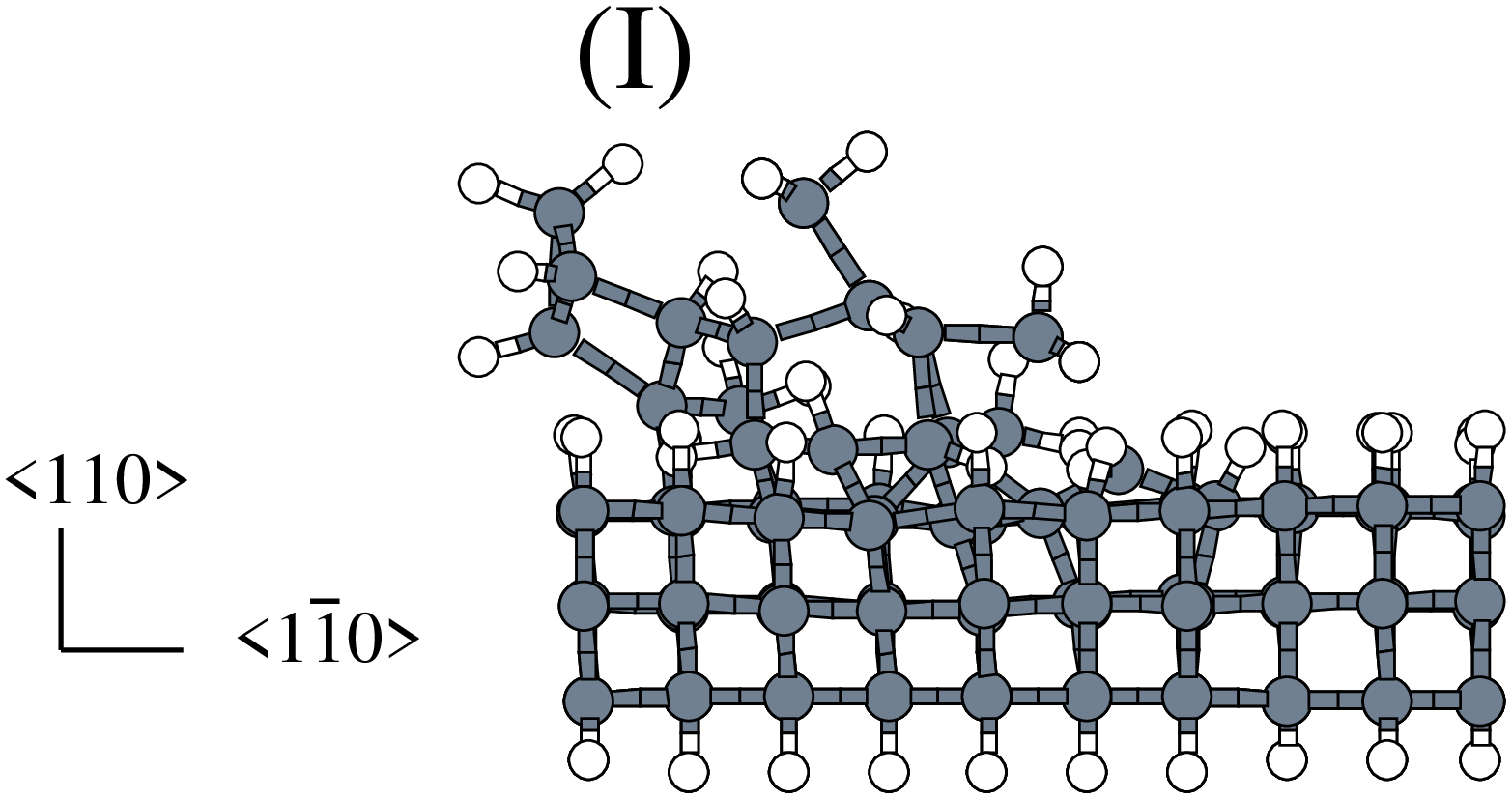}
\vspace*{-2.0cm}
\end{center}                                                           
\caption{Atomic configurations, corresponding to points shown on the force graph (Figure \ref{Forces}), relating to the simulation performed in the hard polishing direction.  The diagrams correspond to tip positions (E) = 1.7 \AA, (F) = 2.4 \AA, (G) = 2.7 \AA, (H) = 4.4 \AA \hspace*{0.1cm} and (I) = 4.9 \AA.}
\label{Hard}                                                           
\end{figure}

Figure \ref{Forces} is of particular interest because it allows us to identify the particular atomic configurations which are responsible for changes in the asperity's degree of resistance to deformation.  Points where the key processes occur have been labelled in the figure and the configurations corresponding to these points are shown in Figures \ref{Soft} \& \ref{Hard}.  Points relating to the soft direction will be discussed first.
 At position (A) both the normal and the retarding forces have been slightly reduced by the loss of a hydrogen from the asperity and the subsequent formation of a C--C bond between the asperity and the tip \cite{R16}.  This rearrangement does not alter the tetrahedral coordination of the atoms in the asperity which persists until the first C--C bond in the asperity is broken (B).  Beyond this point the asperity is compressed, mainly through bond angle deformation, until, between (C) and (D), there is a reconstruction in which the asperity loses its remaining structural strength through the breaking of a six--membered ring.  At this stage the absence of structural rigidity permits the few remaining single C--C bonds, by which the asperity now adheres to the surface, to be broken with relative ease.  At the termination point, the structure of the asperity is neither diamond--like nor graphitic and any polishing residue is likely to be amorphous \cite{R17}.  Throughout the process the deformation remains localized, affecting only the atoms within the asperity itself.  This suggests that the surface left behind will be flat on the atomic scale.  A quantitative estimate of the upper limit on the wear rate was made using numerical data from the simulation, under the assumption that this process is entirely responsible for the experimentally measured normal load \cite{R18}.  This estimate is significantly above the maximum experimentally observed wear rate indicating that the process can easily account for the magnitude of the observed wear.


Turning now to the hard direction (Figure \ref{Hard}), position (E) is
where a carbon in the tip, which had already bonded to a hydrogen atom from
the asperity, breaks this bond, and rebonds to a carbon atom in the asperity,
leaving the hydrogen free to bond to a different carbon in the tip.
As the tip is advanced towards (F) more bonds are formed between the
tip and the asperity, although in supplying the large force on
the tip the asperity has only deformed to a limited extent.  Between
(F) and (G) the force on the tip falls off dramatically and the region of the
asperity in the immediate neighbourhood of the tip reconstructs.  The
effect of the reconstruction is to force the two carbons at the
leading edge of the asperity beneath the tip.  This causes significant
sub--surface damage which may even have been slightly restricted by the
rigid base plane.  The structure after the transformation looks
similar to the structure at the position (E) where the force began to
accumulate.  This suggests that, if the asperity were longer, and
therefore more able to support a large retarding force, the forcing of the
carbon atoms into the sub--surface could be repeated.  At (H) the
remaining asperity carbons are in a highly distorted configuration, primarily due to large bond angle distortions, and the atoms at the back of the asperity yield first
leading to a complete loss of rigidity.  At point (I) the retarding
force has fallen significantly and it is likely that, as in the
soft direction, the remains of the asperity would form an amorphous
residue.  However, in contrast to the soft direction, the deformation
in the hard direction is highly non--local.

\begin{figure}[htbp]                  
\begin{center}
\hspace*{0.0cm}                           
\epsfxsize=8.0cm                      
\epsffile{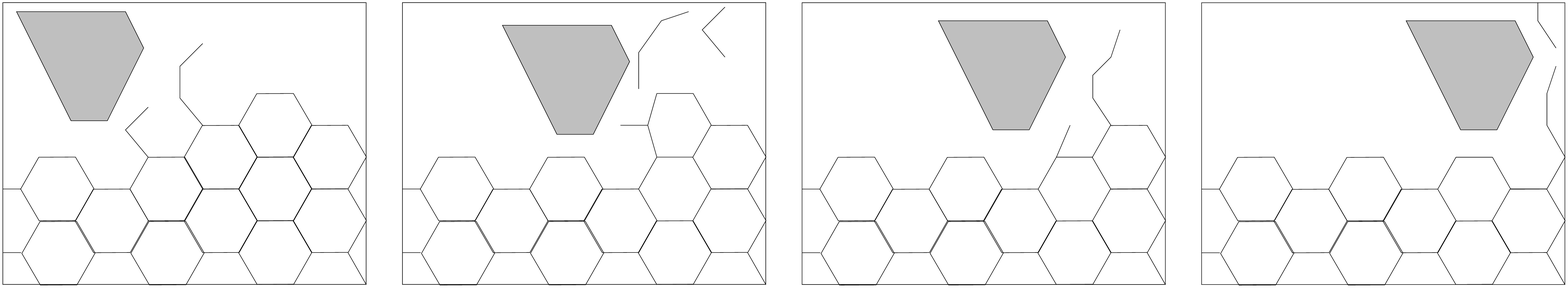}
\vspace*{-0.0cm}   
\end{center}                          
\caption{Carbon skeleton schematic representation of suggested nano--grooving mechanism.  The top figure is for the hard direction and shows the deformation to be non--local.  The lower figure is for the soft direction and shows the mechanism to be local and repeatable.}
\label{Speculate}                     
\end{figure}

The process of removing an asperity may be related to that of forming a nano--groove by making two changes to the system.  Firstly, a change in boundary conditions at the edge of the asperity is required to recreate an ideal surface.  Secondly, the chemical environment beneath the tip must be altered owing to the removal of the preceding asperity.  The effect of making these changes is very different for the two directions.  In the soft direction the change in boundary conditions involves linking the (110) chains to join asperities in adjacent supercells.  A change in the chemical environment beneath the tip will mainly affect the normal force which, as already discussed, is not believed to be important.  The local nature of the deformation during asperity removal suggests that both the mechanism and the forces occurring during nano--groove formation will be similar to those observed in the simulations.  In contrast, in the hard direction the forces occurring during nano--groove formation are expected to be significantly higher than those arising during the removal of an asperity, although the mechanism will be similar.  The change in boundary conditions has the effect of extending the rear of the asperity to infinity.  Consequently it is able to supply much greater resistance to the oncoming tip.  The chemical environment beneath the tip may be significantly altered as the sub--surface is likely to be damaged.  As already mentioned, the process of forcing carbon into the sub--surface may be repeated until either the increasing normal force due to the compressed carbon deflects the tip or bond breaking occurs at an intermediate distance from the tip leading to the removal of carbon.  Figure \ref{Speculate} shows the suggested nano--grooving mechanisms.

                                       
In summary, we propose the mechanical process of nano--grooving as the principal mechanism for wear of the (110) surface of diamond during polishing in the soft direction.  The mechanism proceeds mainly via bond angle deformation, and stress is localized on a few diamond bonds at a time, allowing bond breaking.  The role of the normal force does not appear to be important, the polishing residue is expected to be amorphous and the surface will be left atomically flat.  In the hard direction nano--grooving may play a role in the wear of diamond but the process is significantly more difficult.  Provided that a normal force is present to stabilize the surface the deformation is not local and the stress is distributed over a large number of diamond bonds.  There is extensive subsurface damage and the likely polishing residue would again be amorphous.  We believe that these differences in the atomic mechanisms of nano--groove formation are responsible for the observed anisotropy in the polishing rate.

M.\ R.\ J.\ acknowledges the financial support of the EPSRC.  
R.\ P.\ acknowledges the financial support of the 
CICYT under project PB92--0168.
This work has been partially supported by Acciones Integradas
Hispano--Britanicas (MEC--British Council) under project HB1996--0023.
Computer time was provided on the Cray T3D at EPCC in Edinburgh.

\vspace*{0.4cm}                       
\vspace*{-0.2cm} 

\begin{figure}[htbp]
\vspace*{-1.5cm}                   
\begin{center}                         
\hspace*{-0.4cm} \epsfxsize=5.0cm \epsffile{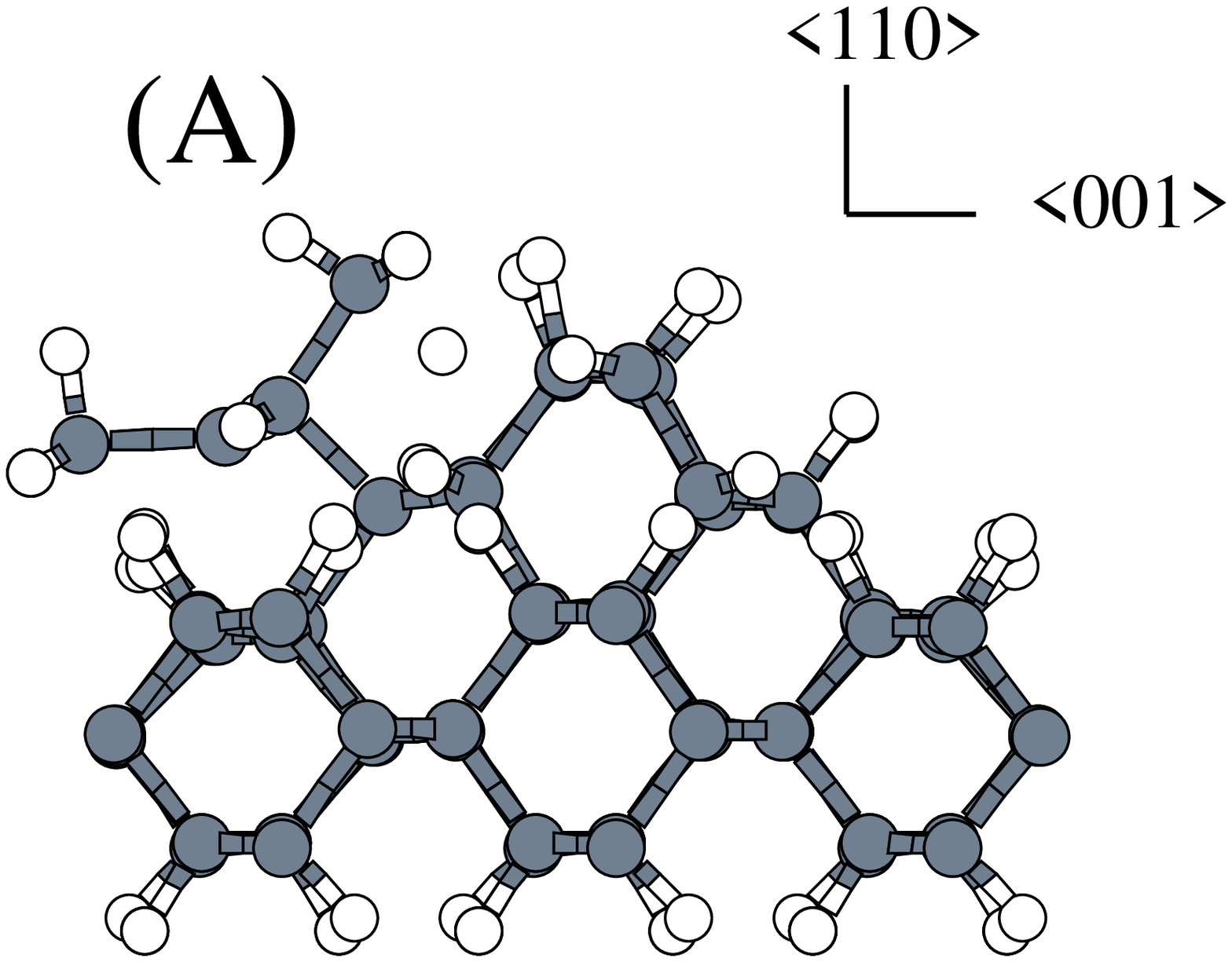} \hspace*{-1.2cm} \epsfxsize=5.0cm \epsffile{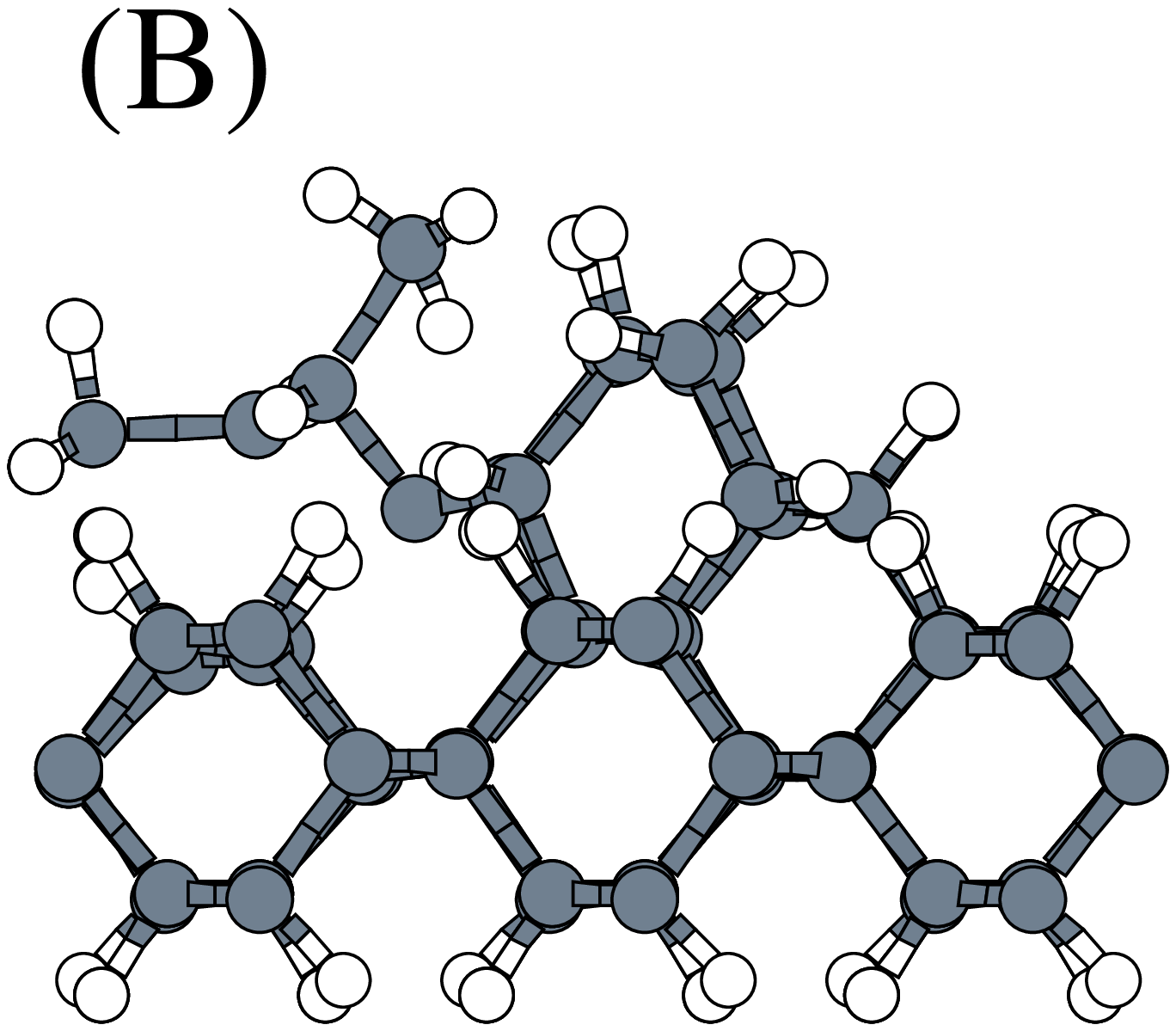}   
\end{center}                           
\end{figure}                                                                  
\vspace*{-4.2cm}                       

\begin{figure}[htbp]                                                   
\vspace*{-2.2cm}
\begin{center}\hspace*{-0.6cm} \epsfxsize=5.0cm \epsffile{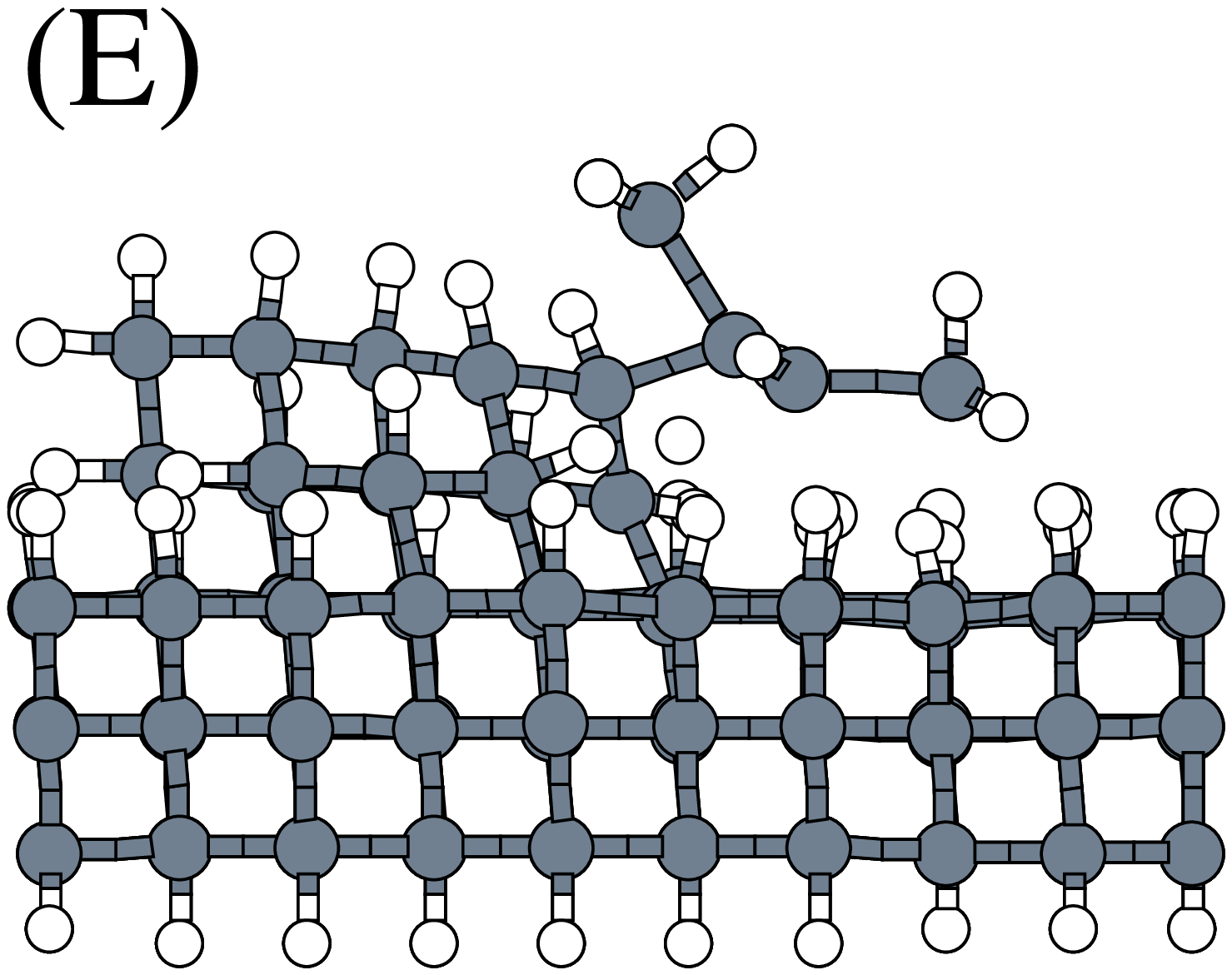} \hspace*{-0.7cm} \epsfxsize=5.0cm \epsffile{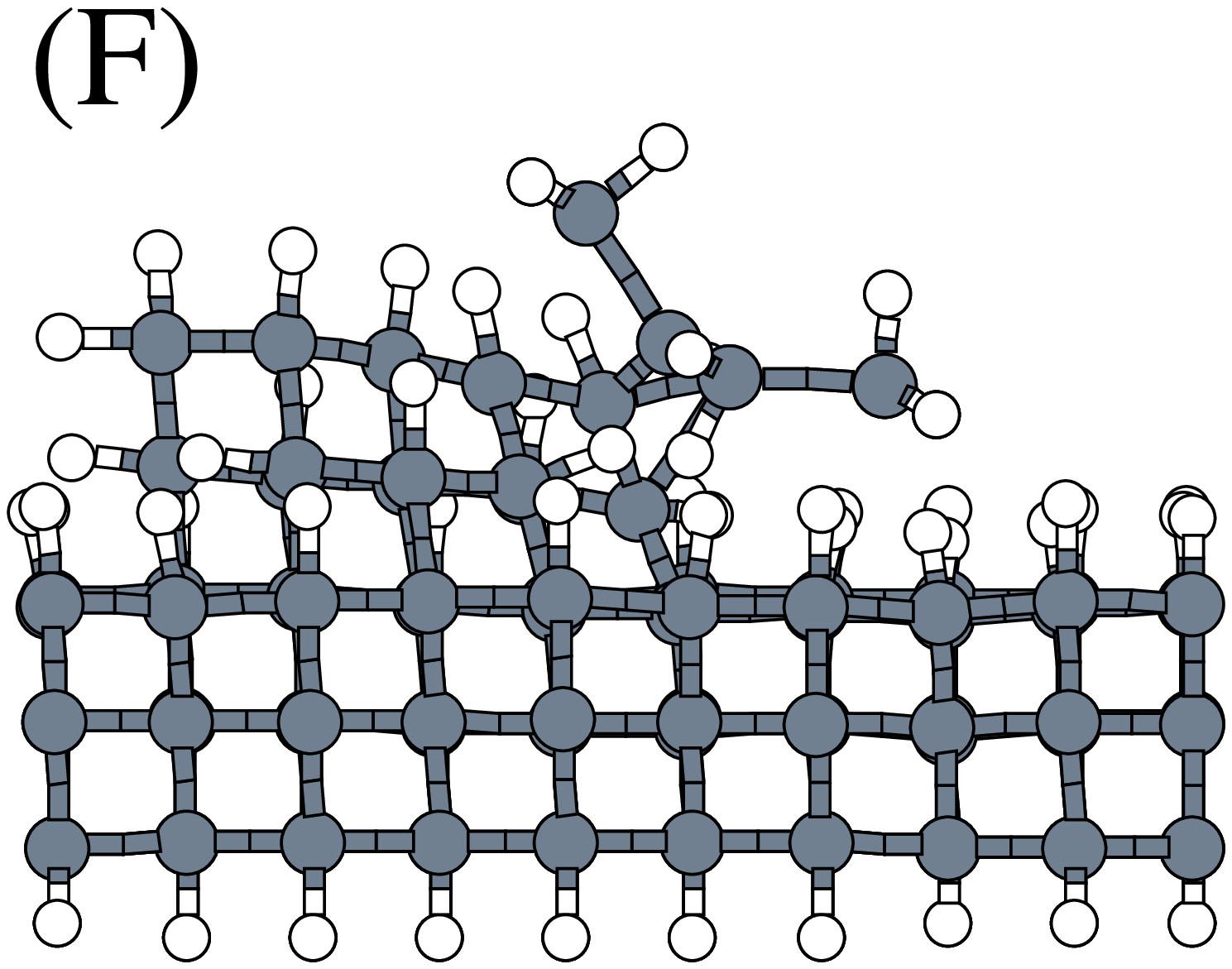} 
\end{center}                                                           
\end{figure}                                                                   
\vspace*{-5.0cm}                                                        
\begin{figure}[htbp]                                                   
\begin{center}\hspace*{-0.6cm} \epsfxsize=5.0cm \epsffile{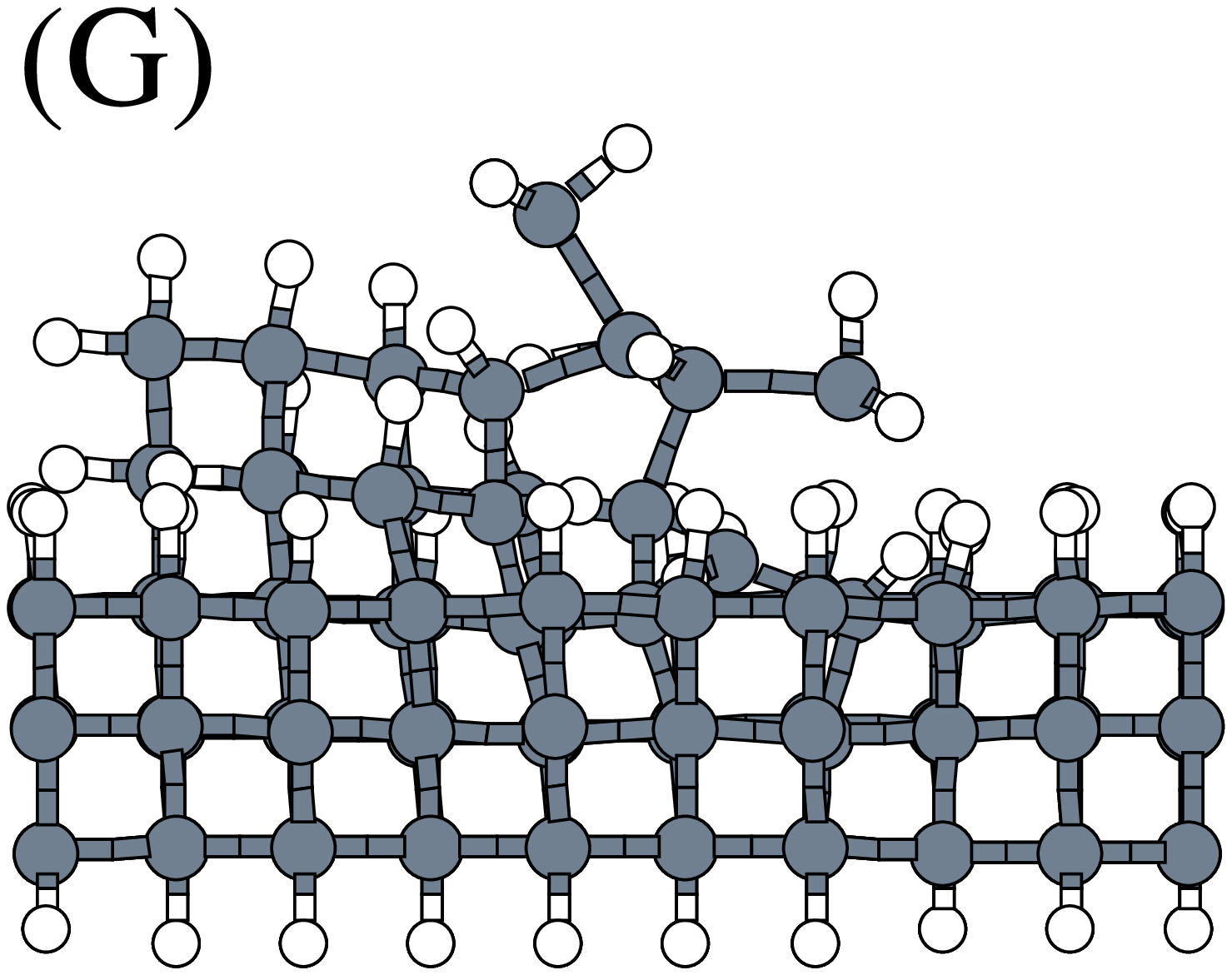} \hspace*{-0.7cm} \epsfxsize=5.0cm \epsffile{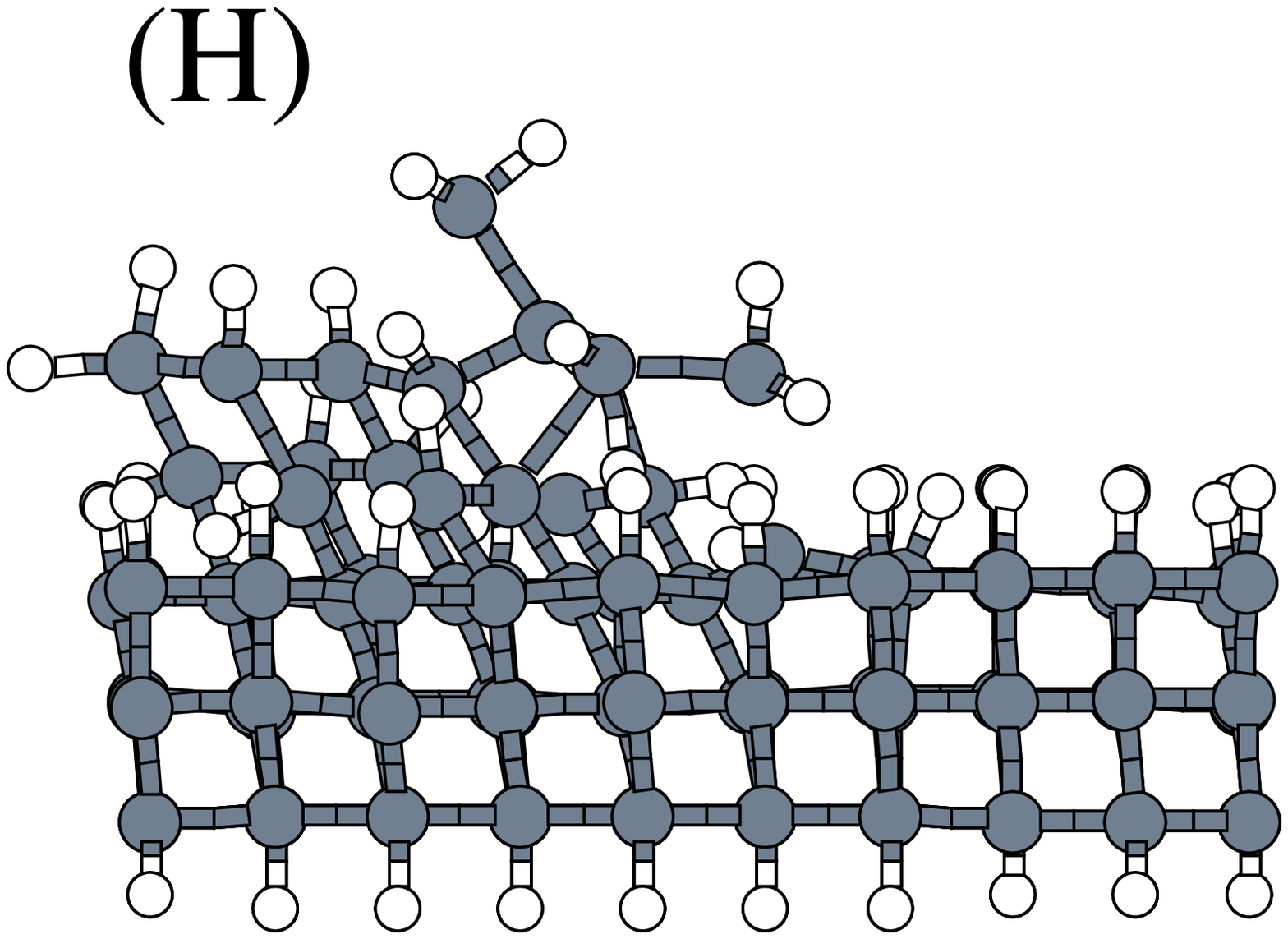} 
\end{center}
\end{figure}
\vspace*{-4.6cm}                                                       

\vspace*{-0.2cm}                         
\begin{figure}[htbp]                     
\begin{center} 
\hspace*{0.0cm}                          
\epsfxsize=8.0cm                         
\epsffile{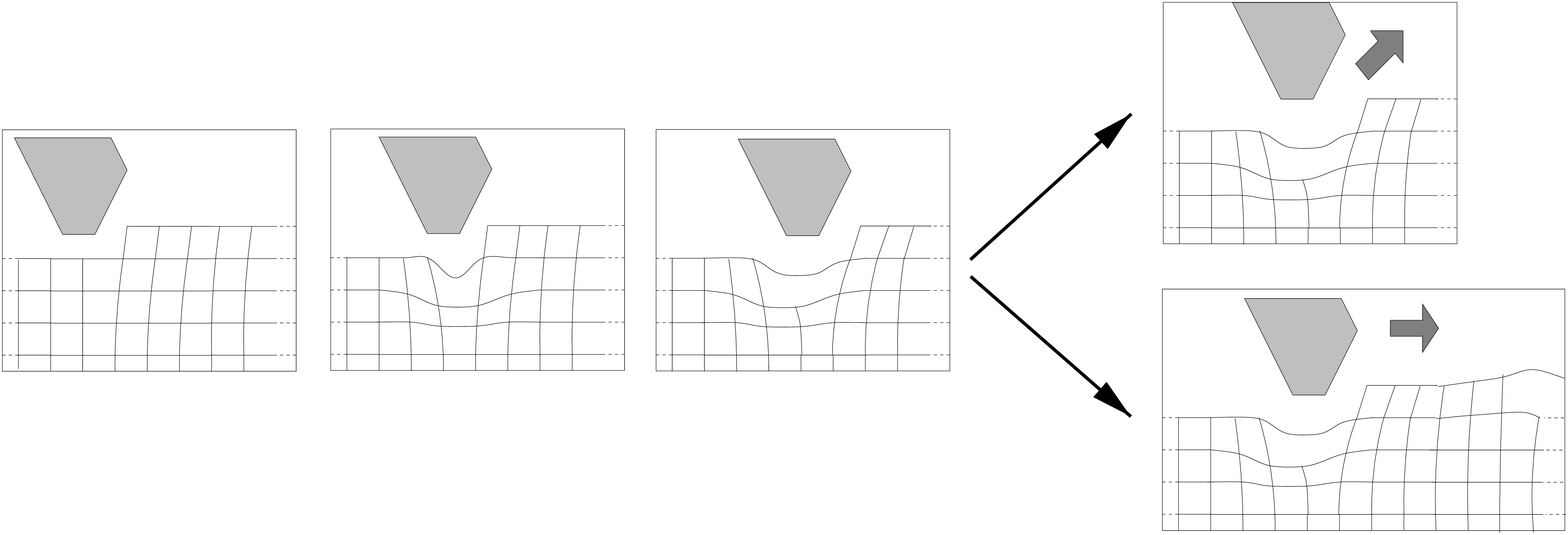} 
\end{center}                                                     
\end{figure}                             
 
\vspace*{-1.0cm}                      
\vspace*{-0.00cm}


\begin{references}
\bibitem{R2} The experimental evidence regarding polishing on the (110) and the (110) surfaces is similar but different from the relatively hard (111).  The (110) was chosen for study as the $90^o$ asymmetry on this surface allows us to use a smaller supercell than would be required for the (100) which has a $45^o$ asymmetry.
\bibitem{R15} An additional plane of carbon atoms was required for the hard direction to avoid chemical interaction between the tip in one supercell and the asperity in the next.
\bibitem{R16} We define a bond to be formed when the relevant atoms are separated by less than the sum of their covalent radii i.e. 1.72 \AA \hspace*{0.05cm} for a C--C bond and 1.22 \AA \hspace*{0.05cm} for a C--H bond.
\bibitem{R17} Beyond this point, the absence of rigidity made the Born--Oppenheimer surface shallow and hence the validity of the adiabatic approximation became questionable and the calculation became expensive to converge.
\end{references}
\end{document}